\documentclass[aps,pra,twocolumn,superscriptaddress]{revtex4-1}
\usepackage{amsmath}
\usepackage{amsfonts,bbold}
\usepackage{amssymb}
\usepackage{enumerate}

\usepackage{graphicx}
\usepackage{subfigure}
\usepackage{natbib}
\usepackage[colorlinks=true,linkcolor=blue,urlcolor=black,citecolor=blue]{hyperref}

\newcommand{\tr}{\mathrm{tr}}

\begin{document}

\title{Geometric decoherence in diffusive open quantum systems}

\author{Da-Wei Luo}

\affiliation{Center for Quantum Science and Engineering and Department of Physics, Stevens Institute of Technology, Hoboken, New Jersey 07030, USA}
\affiliation{Beijing Computational Science Research Center, Beijing 100094, China}

\author{Hai-Qing Lin}
\affiliation{Beijing Computational Science Research Center, Beijing 100094, China}

\author{J. Q. You}
\affiliation{Department of Physics and State Key Laboratory of Modern Optical Instrumentation, Zhejiang University, Hangzhou 310027, China}

\author{Lian-Ao Wu}
\affiliation{Department of Theoretical Physics and History of Science, The Basque Country University (UPV/EHU), PO Box 644, 48080 Bilbao, Spain}
\affiliation{Ikerbasque, Basque Foundation for Science, 48011 Bilbao, Spain}

\author{Rupak Chatterjee}
\affiliation{Center for Quantum Science and Engineering and Department of Physics, Stevens Institute of Technology, Hoboken, New Jersey 07030, USA}

\author{Ting Yu}
\email{Corresponding Author:  Ting.Yu@stevens.edu}
\affiliation{Center for Quantum Science and Engineering and Department of Physics, Stevens Institute of Technology, Hoboken, New Jersey 07030, USA}

\date{\today}
% Oct 25, 2019

\begin{abstract}
Based on a generic quantum open system model, we study the geometric nature of decoherence by defining a complex-valued geometric phase through stochastic pure states describing non-unitary, non-cyclic and non-adiabatic evolutions.  The ensemble average of the complex geometric phases for the pure stochastic states yields a conventional geometric phase together with an amplitude factor.  We show that the decoherence process described by the decaying amplitude can be a geometric quantity independent of the system's dynamics.
It is a remarkable fact that the geometric phase of a quantum system can serve as an ideal realisation of quantum gates due to its robustness against dynamical errors, however, in this paper we show that, for some open quantum systems, a desirable geometric phase may be accompanied by an unwanted robust geometric decoherence factor. Two exactly solvable models are studied to demonstrate that, while the decoherence is a purely dynamical effect for a dephasing two-level model, the decoherence in a dissipative two-level model can be
a geometric process.  Finally,  we show that such a geometric decoherence effect may be eliminated by a non-perturbative control scheme.

\end{abstract}

\maketitle

\section{Introduction}

The dynamics and decoherence processes of quantum open systems have been under intense research~\cite{Breuer2002} in various different fields, most prominently in quantum foundation \cite{zurek,zeh}, quantum optics~\cite{Scully1997a} and quantum information~\cite{Nielsen2000}. The decoherence process is mostly understood as a dynamical process 
due to the coupling to an external environment. As all quantum systems are interacting with their surroundings to some degree, the theory of open system dynamics provides a complete description of the quantum system under consideration~\cite{Carmichael2002,Plenio_Knight_1998}. The environment or bath that the system is embedded in is commonly modelled as a set of  bosons, fermions or spins, and the initial state of the bath along with other details such as the modes' frequencies and coupling strengths can be encoded in a bath correlation function, summarizing the profile of the bath to account for its influences on the system being studied. The quantum open system models can be treated in many different ways, giving rise to different decoherence effects. One of the major motivations of this paper is to identify under what conditions the quantum decoherence may be described as a geometric quantity varying with time and to quantify the geometric decoherence for different types of environmental noises. A primitive approach to quantum open systems is to treat the bath as essentially memory-less, where the dynamics of the system is not influenced by the history of the bath, and is generally known as the Markovian approximation. While it is a somewhat valid approximation for weakly-coupled systems, it has been realized that for more general cases, this Markovian description is inadequate, and fails to capture some very interesting memory effects. To fully describe the open system dynamics, many approaches have been developed to deal with the non-Markovian open systems~\cite{Breuer2002}, such as the path integral approach and the stochastic Schr\"odinger equation approach. A very notable example of the latter is the quantum state diffusion equation~\cite{Diosi1998}, which has been systematically developed to deal with systems linearly coupled to bosonic as well as fermionic baths, without specifying the exact details of the system Hamiltonian or the coupling mechanism. Exact or approximate analytical solutions can be derived for many interesting systems~\cite{Diosi1998,Jing2010,Yu1999a,Zhao2012,Jing2013}. Exact master equations may also be obtained~\cite{Chen2014}, and various numerical methods have also been proposed to deal with arbitrary systems~\cite{Li2014,Luo2015,Suess2014a}.

The geometric phase associated with quantum evolutions~\cite{Berry1984,Pancharatnam1956,F-Wilczek1989a} has offered a lot of insight into many interesting phenomena such as quantum phase transitions~\cite{Carollo2005} and the spin Hall effects~\cite{Bliokh2008a}. The geometric phase is an observable quantity that's dependent only on the path traced out by the state in the parametric manifold, or in the projective Hilbert space, and is independent of other details such as speed of the evolution or the equation of motion, which does not need to be a Schr\"odinger-like equation~\cite{Samuel1988,F-Wilczek1989a,measure_note}. Over the last decades, various extensions to the original adiabatic geometric phase have been proposed, including the non-adiabatic AA phase~\cite{Aharonov1987}, non-cyclic or non-Hermitian case~\cite{Samuel1988,Pati1995,Wu1994a}, as well as non-Abelian systems~\cite{Anandan1988a,Wilczek1984a}, where one may get a matrix-valued extension of the geometric phase, which is under very active scrutiny due to its potential use in holonomic quantum computation~\cite{Zanardi1999,Wu2005a,Pyshkin2016a}. Mathematically, the geometric phase can best be understood via a parallel transport using a connection on a fiber bundle~\cite{F-Wilczek1989a,Samuel1988} or as the holonomy transform~\cite{Simon1983}. The purpose of this paper is to discuss the geometric processes of a quantum open system coupled to a quantum environment modelled by a set of bosons. Our stochastic Schr\"odinger equation based on a general quantum open system allows for a versatile treatment of generalized complex geometric phases in the presence of various environmental noises ranging from non-Markovian to Markov noises beyond the general non-unitary quantum dynamics~\cite{Samuel1988,Pati1995}. Our method can easily identify the decoherence process associated with the open system's evolution being geometric or dynamic.  It should be noted that other types of open systems extensions of the geometric phase are also discussed using either a purification scheme to deal with mixed states~\cite{Tong2004}, Markovian trajectories~\cite{Carollo2003,Buric2009a} or considering adiabatic evolutions~\cite{Sarandy2006}, and a recent extension to non-Markovian systems where the environments have memory effects~\cite{Luo2017}. In this paper, by using stochastic pure states generated by the stochastic Schr\"odinger  equation, we systematically study the geometry of decoherence of open systems with a complex geometric phase~\cite{Berry1990a,Garrison1988a,Aitchison1992a,Cui2012a,Cui2014a}, whose imaginary contribution is interpreted as a decoherence factor~\cite{gp_dephasing_note}. For this purpose, our focus is primarily on the decaying amplitude information. One of the advantages of our method is that we can directly take into account the environment information such as memory times, coupling strengths and correlations between two environments.

The organization of this paper is as follows. In Sec.~II, we define the complex geometric phase of a general open quantum system based on the stochastic pure states governed by a diffusive stochastic Schr\"odinger equation. To see its connection to geometry in a more transparent way, we rewrite the complex phase as a connection one form. 
The ensemble average of the pure state trajectories will yield the desired information on the geometric dynamics of the open quantum systems. In Sec.~III,
we study two exactly solvable models consisting of a two-level system embedded in a non-Markovian multi-mode bosonic bath. It is found that the decoherence of a single two-level pure dephasing model is a purely dynamical effect, namely, the imaginary part of the geometric phase vanishes. However, for the dissipative model, the geometric component describing the decoherence process can be efficiently identified. We show the geometric decaying factor persists in various environmental memory time scales including the Markovian limit. In Sec. IV, we show how to combat the adversary geometric decoherence effect of the bath, where a non-perturbative control known as the Leakage elimination operator (LEO)~\cite{Jing2015a} is utilized to correct the open system trajectory so that the state stays close to the closed system evolution. As a result, we get a geometric phase that's close to the target closed system one, while correcting for the detrimental influence of the environment. We conclude in Sec.~V, while some useful material can be found in Appendix A.

\section{Complex geometric phase for open quantum systems}

Let us consider a generic quantum system embedded in a bosonic bath ($T=0$), with the total Hamiltonian being (setting $\hbar=1$)
\begin{align}
    H=H_s+\sum_k (g_k L ^\dagger b_k+g_k^* L b_k ^\dagger)+\sum_k \omega_k b_k ^\dagger b_k,  \label{totalh}
\end{align}
where $H_s$ is the system Hamiltonian, $L$ is called the Lindblad operator describing the system-bath coupling mechanism, and $b_k$($b_k ^\dagger$) is the annihilation (creation) operator of the $k$-th bath mode. The influences of the bath on the open system is encoded in a bath correlation function $\alpha(t,s)=\sum_k |g_k|^2e^{-i\omega_k(t-s) }$. In the Markovian case, the correlation function becomes a $\delta$ function, but the Markovian approximation may not be a valid choice for many realistic physical systems and can fail to correctly predict the properties of the system under consideration. In the case of decoherence process, the decoherence time may take place on time scales that can be of the same order as the correlation time of the environment, then the standard Markov approximation is not valid anymore.  The general non-Markovian dynamics of the open quantum system may be obtained systematically through a projection onto the coherent state basis $|z \rangle=|z_1 \rangle\otimes|z_2 \rangle\otimes\ldots$ of the bath modes in the interaction picture with respect to the bath. We can then get a stochastic Schr\"odinger equation living in the system Hilbert space known as the non-Markovian quantum state diffusion (NMQSD) equation~\cite{Diosi1998}
\begin{align}
        \partial_t |\psi_{z^*}(t)\rangle &= \left[-iH_s+Lz_t^*-L ^\dagger \bar{O}(t,z^*)\right]|\psi_{z^*}(t) \rangle \nonumber\\
    & = -iH_{\rm eff} |\psi_{z^*}(t) \rangle, \label{qsd}
\end{align}
where $z_t^*=-i\sum_k g_k z_k^* \exp(i \omega_k t)$, $\bar{O}(t,z^*)=\int_0^t \alpha(t,s)\frac{\delta}{\delta z_s^*}$ is the average of the functional derivative weighted by the memory function, and the pure state $|\psi_{z^*}(t) \rangle$ is called a quantum trajectory. Defining an ensemble average $\mathcal{M}(\mathcal{F})=\int d^2z |z|^2 \mathcal{F}/\pi$, we can see that $\mathcal{M}(z_tz_s^*)=\alpha(t,s)$ and the reduced density operator of the system is given by $\rho=\mathcal{M}(| \psi_{z^*}(t) \rangle\langle \psi_{z}(t)|)$. For many systems, the $\bar{O}$ operator can be analytically obtained, giving an exact non-Markovian description of the dynamics of the system. Furthermore, an exact master equation can also be obtained. For general systems, various numerical techniques have been developed to tackle the problem. The stochastic Schr\"odinger equation~\eqref{qsd} dictates the dynamics of the pure state trajectory evolving under an effective non-Hermitian Hamiltonian. Especially in the Markovian regime, this equation describes the conditioned system state when the bath is continuously measured~\cite{Gambetta2002a}.

For the open system described by~\eqref{qsd},  a stochastic complex geometric phase can be defined for the pure state $|\psi_{z^*}(t) \rangle$. It has been known that for an arbitrary pure state $|\psi \rangle$ governed by an effective Hamiltonian $H_{\rm eff}$, there exists an adjoint state $|\tilde{\psi} \rangle$ evolving under $H_{\rm eff}^\dagger$ so that the norm $\langle \tilde{\psi}|\psi \rangle$ is conserved during evolution. One then tracks this pair of states in the projective Hilbert space, which gives, in general, a complex-valued geometric phase. This approach has the advantage of tracking the geometric effects of non-Hermitian evolutions. Since in the NMQSD approach, the reduced density operator is given by the ensemble average of $|\psi _{z^*}\rangle \langle \psi_{z}|$, changes in the norm of the trajectories may be understood as the weight $p_i$ of some normalized pure state decomposition of the density operator $\rho=\sum_i p_i |\psi_i \rangle\langle \psi_i|$. Accordingly, we can track the geometric component of the dissipation by using the complex-valued extension of the geometric phase.

We start by reviewing the complex geometric phase formulation for non-unitary, non-cyclic evolutions. Consider a generic quantum state $|\psi(t) \rangle$ evolving under $H(t)$ (here $H(t)$ is a generic non-Hermitian operator) and its adjoint $|\tilde{\psi}(t) \rangle$ evolving under $H ^\dagger (t)$. One first defines the complex phase between two states $|a \rangle$, $|b \rangle$   with their respective adjoints~\cite{branch_note},
\begin{align}
    \exp[i \varphi]=\sqrt{\frac{\langle \tilde{a}|b \rangle}{\langle \tilde{b}|a \rangle}}. \label{eq_def_phase}
\end{align}
In this setting, the dynamical phase can be written as $\beta_{\rm dyn}=-\int_0^T ds \langle\tilde \psi(s)|H(s)|\psi(s)\rangle$. To prove this, we may remove the dynamical phase to form a horizontal vector $|\psi_h(t)\rangle= \exp[-i \beta_{\rm dyn}]|\psi(t)\rangle$ and its corresponding adjoint $|\tilde\psi_h(t)\rangle= \exp[-i \beta_{\rm dyn}^*]|\tilde \psi(t)\rangle$. It can be checked that the parallel transport condition $\langle\tilde\psi_h(t)|\partial_t|\psi_h(t)\rangle$ is satisfied so that the generalized phase it acquires is purely geometric. Formally, one may write the geometric phase as the total phase minus the dynamical phase,
\begin{align}
    \beta&=\beta_{\rm tot}-\beta_{\rm dyn} \nonumber\\
    &=-i \ln\left[\sqrt{\frac{\langle \tilde \psi(0)|\psi(t) \rangle}{\langle \tilde \psi(t)| \psi(0) \rangle}}\right]+i\int_0^t ds \langle\tilde \psi(s)|\partial_s|\psi(s) \rangle.\label{eq_gp_cal}
\end{align}
Based on our stochastic Schr\"odinger equation approach, one can identify $|\psi(t) \rangle=|\psi_{z^*}(t) \rangle$ and the corresponding adjoint $|\tilde \psi(t) \rangle=|\tilde \psi_{z}(t) \rangle$ evolving under the effective Hamiltonian $H_{\rm eff}^\dagger =H_s-iL ^\dagger z_t+i \bar{O} ^\dagger (t,z)L$ to obtain the complex-valued geometric phase for a single quantum trajectory $|\psi _{z^*}\rangle$. Following~\cite{Pati1995} to define a generalized reference section on the appropriate bundle
\begin{align}
    |\chi(t)\rangle=\sqrt{\frac{\langle \tilde \psi(t)|\psi(0) \rangle}{\langle \tilde \psi(0)| \psi(t) \rangle}}|\psi(t)\rangle\equiv\exp[-i \varphi(t)]|\psi(t)\rangle
\end{align}
This reference section satisfies the following conditions
\begin{enumerate}[(i)]
\item The initial wave function $|\psi(0)\rangle$ and the initial reference section coincide.
\item $\pi[|\psi(t) \rangle]=\pi[|\chi(t) \rangle]$, where $\pi[\ldots]$ is the map to the projective Hilbert space $\mathcal{P}$. This means that both wave functions project to the same curve in $\mathcal{P}$.
\item $|\chi(t) \rangle$ stays in phase with with $|\chi(0) \rangle$ at all times, where the phase is defined under Eq.~\eqref{eq_def_phase}.
\end{enumerate}
With a corresponding adjoint reference section
\begin{align}
    \langle\tilde\chi(t)|=\langle \tilde \psi(t)| \sqrt{\frac{\langle \tilde \psi(0)|\psi(t) \rangle}{\langle \tilde \psi(t)| \psi(0) \rangle}},
\end{align}
the complex non-cyclic geometric phase may now be given by a integral of a connection one-form
\begin{align}
    \beta=i\int \langle\tilde\chi(s)|\partial_s|\chi(s) \rangle ds.\label{eq_1form}
\end{align}

It can also be proved that the reference section and hence the complex geometric phase is gauge-invariant under a  transformation $|\psi(t) \rangle \rightarrow \exp[i \theta(t)]|\psi(t)\rangle$, where $\theta(t)$ may be complex. This definition does not need an explicitly closed curve in the projective Hilbert space, but one can still join both ends with a geodesic to form a closed curve, so that one can formally write the geometric phase as a surface integral over a solid angle to highlight its geometric nature. To see this is the case,  consider the covariant derivative $D_s=\partial_s-A_s$, where the connection is given by $A_s=\langle \tilde \varphi(s) |\partial_s|\varphi(s) \rangle$. By letting $|u' \rangle=D_s |\varphi(s) \rangle$, one has an inner product $\langle \tilde u'|u' \rangle$ that is gauge-invariant under $|\psi(t) \rangle \rightarrow \exp[i \theta(t)]|\psi(t)\rangle$. This gives a metric on the projective space $dl^2=\langle \tilde u'|u' \rangle ds^2$, where $dl^2$ is the square distance between points $\pi[|\varphi(s) \rangle]$ and $\pi[|\varphi(s+ds) \rangle]$. Using a variation procedure, one can derive the following geodesic equation (See Appendix~\ref{sec_appa})
\begin{align}
    D_s |u' \rangle=D_s^2|\varphi \rangle=[\partial^2_s-\partial_sA_s-A_s\partial_s+A_s^2]|\varphi \rangle=0,\label{eq_geodes}
\end{align}
Formally, a similar one exists for the adjoint. Using the parallel transport condition and the gauge-invariance, one can show that the connection one-form Eq.~\eqref{eq_1form} disappears along this geodesic. It is worth pointing out that this set of NMQSD equations are invariant under the gauge transformation $L \rightarrow L e^{i \varphi}$~\cite{Buric2009a, gauge_note}, since the phase factor can be absorbed into the coupling strength $g_k \rightarrow g_k e^{i \varphi}$. The noise in the NMQSD equation transforms accordingly as $z_t^* \rightarrow w_t^* = e^{i \varphi} z_t^*$, and the $\bar{O}$ operator becomes the functional derivative with respect to the transformed noise $w_t^*$. Also note that the ensemble average $\mathcal{M}[w_t w_s^*]=\mathcal{M}[z_t z_s^*]=\alpha(t,s)$ is only dependent on the norm $|g_k|^2$ and is gauge-invariant under such transform, meaning $w_t$ and $z_t$ represent the same set of stochastic processes. Therefore, the transformed equation represents the same set of equations as Eq.~\eqref{qsd}, and the geometric phase we obtain from them inherits this gauge invariance. The same argument also holds for the model with multiple Lindblad operators $L_m$, where the phase transformation is given by $L'_m=L_m e^{i \varphi_m}$, ($m  =1, 2, \ldots, N$)~\cite{percival}.

%%%%%%%%%
\section{Examples}
%%%%%%%%%%%

\begin{figure}
    \centering
    \subfigure[]{\includegraphics[scale=.8]{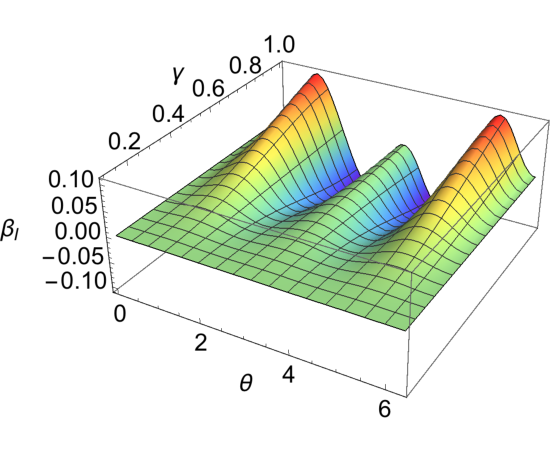}}
    \subfigure[]{\includegraphics[scale=.8]{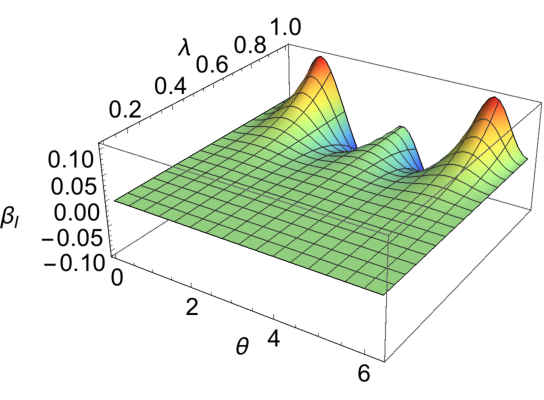}}
    \caption{(Color online)  The imaginary part of the geometric phase $\beta_I$ as a function of the initial state condition $\theta$ and (a) the non-Markovian parameter $\gamma$ of the bath correlation function with $\lambda=1$, (b) the coupling strength $\lambda$ with $\gamma=1$, at $t=2\pi/\omega$. Oscillatory behavior can be observed as $\theta$ changes when $\gamma$ or $\lambda$ are big enough. With small $\gamma$ or $\lambda$, the imaginary part can be very close to zero.}
    \label{fig_tg_tl}
\end{figure}

\begin{figure}
    \centering
    \includegraphics[scale=.7]{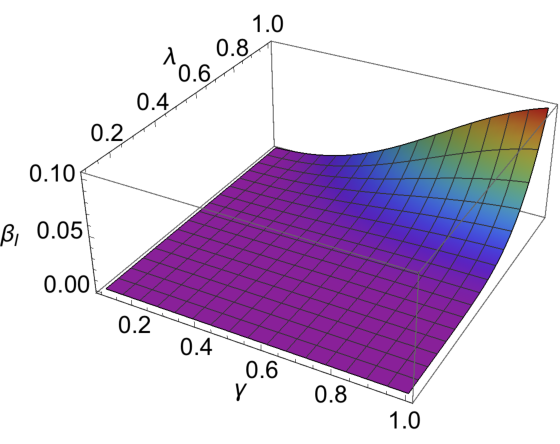}
    \caption{(Color online)  The imaginary part of the geometric phase $\beta_I$ as a function of the coupling strength $\lambda$ and  the non-Markovian parameter $\gamma$ of the bath correlation function, with $\theta=1$ and $t=2\pi/\omega$. It can be readily seen that when $\gamma$ or $\lambda$ are small, the imaginary part of the complex geometric phase is very close to zero, meaning at this region the dissipation is mainly dynamical rather than geometrical in nature.}
    \label{fig_lg}
\end{figure}

After introducing a complex geometric phase for a generic quantum open system~\eqref{eq_gp_cal}, the imaginary part of this complex phase factor will give rise to a geometric amplitude, which is expected to describe quantum decoherence process. By definition, this amplitude factor is purely geometric. We now illustrate the complex-valued geometric phase under the NMQSD to highlight the geometry of the decoherence process. Consider an exactly solvable model, the two-level dissipative model with system Hamiltonian $H_s=\frac{\omega}{2} \sigma_z$ and coupling operator $L=\lambda \sigma_-$. This model is exactly solvable~\cite{Diosi1998,Yu1999a}. Choosing a bosonic bath with a Lorentz spectrum,  the bath correlation function is given by an exponential function,
\begin{align}
    \alpha(t,s)=\frac{\gamma \Gamma}{2}\exp \left[-\gamma|t-s|\right]\exp[-i\omega_0(t-s)],
\end{align}
and in this case, the $\bar{O}$ operator is given by $\bar{O}(t)=F(t)\sigma_-$ where $\dot{F}(t)=\frac{\gamma \lambda}{2}-[\gamma +i (\omega_0-\omega)] F(t)+\lambda F(t)^2$, with $\Gamma = 1$. The initial state here is characterized by the Bloch angle $\theta$, $|\varphi\rangle=[\cos \theta/2,\sin \theta/2]^T$. The total phase, after taking the ensemble average, is given by
\begin{widetext}
\begin{align}
     \bar{\beta}_{\rm tot}=\mathcal{M}\left[-i \ln \sqrt{\frac{\langle \tilde \psi(0)|\psi(t) \rangle}{\langle \tilde \psi(t)|\psi(0) \rangle}}\;\right]=-\frac{i}{2}  \ln \left[\frac{g(t) \left(g(t) (\cos \theta +1)- (\cos \theta -1) e^{i  \omega t}\right)}{ -g(t) (\cos \theta -1)+ (\cos \theta +1) e^{i  \omega t}}\right],
\end{align}
\end{widetext}
where $g(t)=\exp \left[-\lambda \int_0^t F(s)ds\right]$~\cite{branch_note}. The averaged dynamical contribution is given by
\begin{align}
    \bar\beta_{\rm dyn}(t)&=-\int_0^t ds \left[\frac{\omega}{2}   \cos \theta - \frac{i \lambda  F(s) }{2} \left(\cos \theta +1\right)\right].
\end{align}
The geometric phase can then be readily calculated. In Fig.~\ref{fig_tg_tl} we show the imaginary part of the geometric phase as a function of the initial  state parameter $\theta$ and the bath memory parameter $\gamma$ as well as the coupling strength $\lambda$. Here the oscillatory behavior of the phase is beginning to show itself when $\gamma$ or $\lambda$ is not too small. However, in the regime where $\lambda$ or $\gamma$ is indeed small, then the imaginary geometric phase becomes invisible indicating that the dissipation in the regime is dictated by a dynamical effect. To get a clearer picture of this oscillatory behavior, we plot the imaginary geometric phase as a function of the coupling strength $\lambda$ and the memory parameter $\gamma$ in Fig.~\ref{fig_lg} Then one easily sees that there exists a region with small $\lambda$ or $\gamma$ and the imaginary part (amplitude) is small. Moreover, we can expand the geometric phase in terms of the power of $\lambda$ and see that (choosing $\omega_0=0$ and $\omega=1$)
% \begin{widetext}
\begin{align}
    \bar\beta\approx& \frac{t}{2} \cos \theta +\arg \left(\cos \frac{t}{2}-i \cos \theta  \sin \frac{t}{2}\right)\nonumber\\
    &-\frac{i \gamma   \left(e^{i t}-1\right)^2 \sin ^2\frac{\theta }{2} \cos \theta  (\cos \theta +1) e^{-\gamma  t} }{8 (\gamma -i)^2 \left(\cos ^2\theta  \sin ^2\frac{t}{2}+\cos ^2\frac{t}{2}\right)}\times \nonumber \\
    &\left[1+e^{(\gamma -i) t} (\gamma t-i t-1)\right] \lambda ^2+O\left(\lambda ^4\right).
\end{align}
% \end{widetext}
The first two terms are clearly independent of $\lambda$ corresponding to the closed quantum system case. The terms containing  $O(\lambda)$  do not exist and the first-order term is represented by $O(\lambda^2)$. We may also expand the geometric phase in the powers of $\gamma$. In addition to the  similar closed system terms, the first order reads
% \begin{widetext}
\begin{align}
    -&\frac{2  \lambda ^2 \sin ^2\frac{t}{2} (t-\sin t+i \cos t-i)}{\cos 2 \theta +2 \sin ^2\theta  \cos t+3}\times \nonumber\\
    &\sin ^2\frac{\theta }{2} \cos \theta  (\cos \theta +1)  \gamma+O\left(\gamma ^2\right).
\end{align}
% \end{widetext}

In the Markov limit $\gamma \rightarrow \infty$, we recover the standard Markovian case, $F(t)=\lambda/2$, and the complex geometric phase becomes
\begin{align}
    \bar\beta_{\rm M}=&\pi-\frac{1}{2} i \bigg[-2 \tanh ^{-1}\left(\cos \theta  \tanh \frac{\pi  \lambda ^2}{2}\right) \nonumber\\
    &+\pi \cos \theta \left(\lambda ^2+2 i\right)  \bigg]
\end{align} 
at $t=2\pi/\omega$, which can generally be complex. When the coupling strength $\lambda\sim 0$, we return to the closed system case as expected. Indeed, the imaginary part of $\beta_{\rm M}$ is on the order of $\lambda^2$. 

On the other hand, one can consider an analytically solvable example of a two-level system in a pure dephasing environment,  that is, the total Hamiltonian (\ref{totalh}) contains the system Hamiltonian $H_s=\omega \sigma_z/2$ and the coupling operator $L=\lambda \sigma_z$. The $\bar{O}$ operator for this model is just $\int_0^t \alpha(t,s)ds L$. Choosing an initial state characterized by the Bloch angle $\theta$, $|\varphi\rangle=[\cos \theta/2,\sin \theta/2]^T$, one can analytically solve the NMQSD equation~\eqref{qsd} and its dual adjoint evolving under $H_{\rm eff} ^\dagger=H_s-iL ^\dagger z_t+i \bar{O}(t) ^\dagger L$, and use Eq.~\eqref{eq_gp_cal} to obtain the phase. The total phase for a particular trajectory in this model is explicitly given by
%\begin{widetext}
\begin{align}
    \beta_{\rm tot}=&-i \ln\left[\sqrt{\frac{\langle \tilde \psi_{z^*}(0)|\psi_{z^*}(t) \rangle}{\langle \tilde \psi_{z^*}(t)| \psi_{z^*}(0) \rangle}}\right] \nonumber\\
    =&i\lambda ^2 \int_0^t A(s) \, ds \nonumber\\
    & -\frac{1}{2} i \ln \left[\frac{(\cos \theta +1) e^{2 \lambda  \int_0^t z_s^* \, ds}-(\cos \theta -1) e^{i \omega t }}{-(\cos \theta -1) e^{2 \lambda  \int_0^t z_s^* \, ds}+(\cos \theta +1) e^{i \omega t }}\right],
\end{align}
%\end{widetext}
where $A(t)=\int_0^t \alpha(t,s)ds$~\cite{branch_note}. Due to the existence of $z_s^*$ terms in both the numerator and denominator in the second term, one can not directly use the Novikov theorem to compute the ensemble mean over the noise. Noticing that in the second term only $\lambda$ accompanies the noise terms, we can expand it in powers of $\lambda$, where it is found that no $z_tz_s^*$ pair exists, and the mean is given by
\begin{align}
    \bar\beta_{\rm tot}=i\lambda ^2 \int_0^t A(s) \, ds+\arctan\left[\cos \frac{ \omega t}{2}, -\cos \theta  \sin \frac{ \omega t}{2}\right] \label{1sz_tot}
\end{align}
where $\arctan[x,y]=\arctan(y/x)$, which takes into account the quadrant of point $(x,y)$. The dynamic part may be calculated in a similar fashion, 
\begin{align}
   \bar \beta_{\rm dyn}=i \lambda^2 \int_0^tds  A(s)-\frac{1}{2} \omega t  \cos \theta
\end{align}
Interestingly, it is found that the ensemble average geometric phase is the same as the closed system case,
\begin{align}
    \bar \beta=\frac{1}{2} \omega t  \cos \theta+\arctan\left[\cos \frac{ \omega t}{2}, -\cos \theta  \sin \frac{ \omega t}{2}\right].\label{eq_close2}
\end{align}
At $t=2\pi/\omega$, $\beta=\pi \left(\cos \theta+1\right)$. In this case, the geometric phase is robust against dephasing effects. Under this definition of the geometric phase, the pure dephasing process for the two-level system is a dynamical effect.

\section{Leakage control of decoherence}

\begin{figure}[t]
    \centering
    \includegraphics[scale=.7]{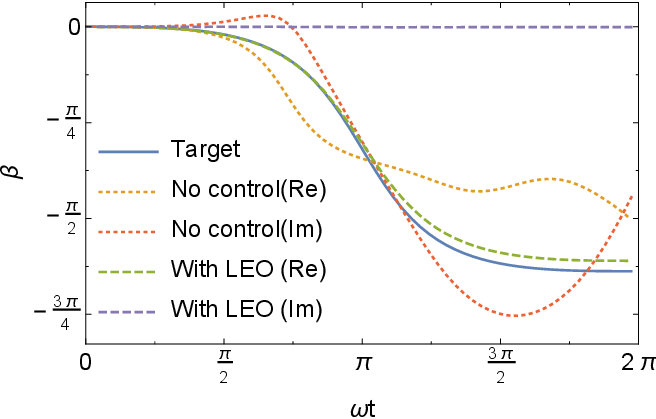}
    \caption{(Color online) The complex geometric phase for a three-level system as a function of time, where the target (solid blue line) corresponds to the closed system case, the dotted orange line and the red dash-dotted line represented the real and imaginary part of the geometric phase, respectively. The green (purple) dashed line is the real (imaginary) part of the geometric phase obtained after the LEO control is applied, where the imaginary part of the geometric phase in the controlled system is close to zero, $|\beta_I|<0.005$. The open system parameters are $\gamma=0.3$ and $\lambda=1$, with an initial state parameter $\theta=1.2$.}
    \label{fig_leo}
\end{figure}

To see how to combat the detrimental effect of the bath on the system's dynamics, we note that a wide range of control strategies have been developed. In the dynamical decoupling control~\cite{Viola1998a},  when the control pulses are applied to the system of interest, it is assumed that the external pulses can be treated perturbatively such that the open system evolves under the pulse Hamiltonian alone during the pulse's active time.  In a non-Markovian open system setting, this assumption may not be accurate. Therefore, here we may apply a non-perturbative control strategy~\cite{Jing2015a} which treats the additional control consistently, and has the additional advantage that only the time integral of the control plays a significant role, making it more resilient against control fluctuations. As an example, we consider a three-level system $H_0=\mathrm{diag} (\omega/2,-\omega/2,0)$, with the  system-bath coupling operator $L=\lambda \left[|3 \rangle\langle 1|+ |3 \rangle\langle 2|\right]$. The corresponding LEO control Hamiltonian is given by $R(t)=c(t)\mathrm{diag} (1,1,0)$, where $c(t)$ is the control function, so the total system Hamiltonian together with the control part is given by $H_s(t)=H_0+R(t)$. This model can be exactly solved by using the NMQSD approach~\cite{Jing2015a}, where the $\bar{O}(t)=F_1(t) |3 \rangle\langle 1|+F_2(t) |3 \rangle\langle 2|$ operator is noise-independent, with
\begin{widetext}
\begin{align}
    \partial_t F_1(t)=\alpha(0) \lambda+\frac{F_1(t)}{2} \left[2 \lambda (F_1(t)+F_2(t))-2 \gamma+i \omega+2ic(t)\right],\nonumber\\
    \partial_t F_2(t)=\alpha(0) \lambda+\frac{F_2(t)}{2} \left[2 \lambda (F_1(t)+F_2(t))-2 \gamma-i \omega+2ic(t)\right].
\end{align}
\end{widetext}
While an analytical solution for the trajectory is absent, there is a convenient way to do the ensemble average for this example. For this model with an initial system state of $[\cos \theta/2,\sin \theta/2,0]^T$, we find that the first two elements of the state vector and its adjoint are all noise-independent. Therefore, the total phase is noise-independent in this case. However, the dynamical phase is noise-dependent. To analytically treat its ensemble average, we define an operator $P(t)=| \psi_{z^*} (t)\rangle \langle \tilde{\psi}_{z^*}(t)|$. Since the adjoint state $|\tilde{\psi}_{z}(t) \rangle$ follows $H ^\dagger$, $P(t)$ is a function of $z_t^*$ only. Using the Novikov theorem, one can show that the ensemble average of the dynamical phase can be written as
\begin{align}
    \mathcal{M}[\phi_d]&=-\int_0^tds \mathcal{M} \left[\langle\tilde{\psi}_{z^*}(s)|H_{\rm eff}(s)|\psi_{z^*}(s) \rangle\right] \nonumber\\
    &=-\int_0^tds \tr \left[H_s(s) \tilde\rho(s)-iL ^\dagger \bar{O}(s) \tilde\rho(s)\right],
\end{align}
where $\tilde\rho(t)= \mathcal{M}[P(t)]$ follows
\begin{align}
    \partial_t \tilde\rho(t)=-i \left[H_s(t),\tilde\rho(t)\right]-\left[L ^\dagger \bar{O}(t),\tilde\rho(t)\right].
\end{align}
Choosing a sine-function control field $c(t)=c_x(1+\sin \Omega_c t)$ with $c_x=10$, $\Omega_c=50$, in Fig.~\ref{fig_leo}, we plot the geometric phases as functions of time for the both LEO controlled and uncontrolled systems. As a comparison, the geometric phase for the closed system described by $H_0$ is also plotted.   It can be seen from Fig.~\ref{fig_leo} that the environmental effect adversely affects the geometric phase and makes it deviate from the target determined by the closed system's evolution. The non-zero imaginary part gives rise to the so-called geometric decoherence. Under the LEO control, the open system can be brought back to a desirable quantum state that is close to the target state, thereby restoring the geometric phase and at the same time suppressing the geometric decoherence by bringing the imaginary part of the generalized geometric phase close to zero.

\section{Conclusion and discussion}

In conclusion, we study the geometric decoherence for a set of general quantum open system embedded in a bosonic bath through defining a stochastic complex-valued extension of the geometric phase for a non-unitary quantum system. The quantum open system coupled to a multi-mode bosonic bath is shown to be governed by a stochastic Schr\"odinger equation known as the NMQSD equation, the reduced density operator for the open system may be recovered by the ensemble average of the pure state trajectories, which evolve under a non-Hermitian Hamiltonian. We associate the imaginary part of the complex geometric phase of the open system with the dissipation and dephasing induced by the bath. By using the two-level systems as illustrative examples, we show that the decoherence of the dephasing model is purely a dynamical effect. For the dissipative two-level system, however, there exists a non-zero imaginary part of the geometric phase, indicating the onset of a geometric component to the dissipation of the open quantum system. As expected, when the coupling strength $\lambda$ is small, the imaginary part is shown to be small consistent with our general understanding of the decoherence processes. 
Our approach allows a more general discussion on the geometric decoherence of open quantum systems across the parameter ranges including both Markov and non-Markovian regimes. We also show that a non-perturbative control scheme can be employed to correct the open system trajectory. As an example, we use a three-level system as an example to show the Leakage Elimination Operator control can correct the system state so that the state can stay close to the closed system, thus generating a robust geometric phase without the side-effect created by the geometric decoherence.

It should also be noted that akin to the real-valued geometric phases in open systems~\cite{Tong2004,Carollo2003,Luo2017}, one may define geometric entities differently, and obtain different geometric phases. For example, a specific model considered in~\cite{gp_dephasing_note} using Markov master equations in an adiabatic evolution setting allows one to define a geometric phase from one density matrix element whose evolution equation is known. In this definition, an imaginary geometric contribution inversely proportional to $T_2$ is obtained, whereas with our general approach the pure dephasing model includes no imaginary part.

The method presented in this paper may be extended to other open quantum systems that interact with fermionic baths via a fermionic NMQSD equation~\cite{Zhao2012a,Shi2013a,Chen2013a}. Another promising direction lies in the application of our approach to study the quantum phase transition in a non-Markovian open system setting~\cite{qpt_note}, where it will be very interesting to study the connection of the geometric or topological aspects with interesting phenomenon in quantum phase transitions such as symmetry-breaking and the thermalization process. We leave these interesting open problems to further investigations.

\begin{acknowledgments}
This work is supported by NSF PHY-0925174. J. Q. Y is supported by the National Natural Science Foundation of China Grant No. 11774022 and the National Key Research and Development Program of China Grant No. 2016YFA0301200. L. -A. Wu is supported by the Basque Country Government (Grant No. IT986-16) and PGC2018-101355-B-I00 (MCIU/AEI/FEDER,UE). H.-Q. Lin, J.Q. You, and D.-W. Luo also acknowledges support from NSAF U1530401.
\end{acknowledgments}

%%%%%%%%%%%%%%%%%%%%%%%%%%%%%%%%%%%%%
\onecolumngrid
\appendix

\section{Proof on the geodesic equation}\label{sec_appa}
Let $|u' \rangle=D_s |\varphi \rangle$ and denoting $|u \rangle=\partial_s|\varphi  \rangle$, we have $|u' \rangle=|u \rangle-\langle \tilde \varphi|u \rangle |\varphi \rangle$. From the normalization condition, we also have $\langle \tilde \varphi|u \rangle=-\langle \tilde u|\varphi \rangle$, and $\delta \langle \tilde \varphi|\varphi \rangle=0=\langle \delta \tilde \varphi|\varphi \rangle+ \langle \tilde \varphi| \delta \varphi \rangle$. Therefore, the variation of $\int \langle \tilde u'|u' \rangle dl$ is given by

\begin{align}
    \Delta &=\delta\int{\left[\langle \tilde u|u \rangle-\langle \tilde u|\varphi \rangle \langle \tilde \varphi|u \rangle\right] }dl\\
    &=\int dl\left[\langle \delta \tilde u|u \rangle+\langle \tilde u| \delta u \rangle\right.\\
    &\left.-\langle \delta\tilde u|\varphi \rangle \langle \tilde \varphi|u \rangle-\langle \tilde u|\delta\varphi \rangle \langle \tilde \varphi|u \rangle-\langle \tilde u|\varphi \rangle \langle\delta \tilde \varphi|u \rangle-\langle \tilde u|\varphi \rangle \langle \tilde \varphi|\delta u \rangle\right].
\end{align}
After some algebra, it can be shown that
\begin{align}
    \Delta
    &=\int dl\left[\langle \delta \tilde u| u' \rangle+ \langle \tilde \varphi| u \rangle \langle\delta \tilde \varphi|u' \rangle\right]
    +\int dl\left[\langle\tilde u' | \delta u \rangle-\langle \tilde u'|\delta\varphi \rangle \langle \tilde \varphi|u \rangle\right]\\
\end{align}
Integrate by parts and let the variation $\delta \varphi$ be zero at both ends, we have
\begin{align}
    \int dl \langle \delta \tilde{\varphi}| \left[|  \partial_l u' \rangle-\langle \tilde \varphi| u \rangle |u' \rangle\right]=0,
\end{align}
which needs to hold for all variational $\delta \tilde{\varphi}$. This then gives us the geodesic equation in Eq.~\eqref{eq_geodes}.

%%%%%%%%%%%%%%%%
\twocolumngrid

% \bibliographystyle{prs}
% \bibliography{dualqsd}

\begin{thebibliography}{10}
\bibitem{branch_note}
To choose a branch for the square root, one may write down the numerator and denominator in polar coordinates, $ze^{i \varphi_z}=\sqrt{\frac{xe^{i \varphi_x}}{y e^{i \varphi_y}}}$, where $x(y,z)$ and $\varphi_{x(y,z)}\in R$. Furthermore, make the following restrictions: $\varphi_x\in (-\pi, \pi]$ and $\varphi_y\in[-\pi, \pi)$. Therefore, $\varphi_z\in (-\pi,\pi]$ and the resultant total phase reduces to the closed system case when the evolution is unitary. Note that a different choice of the branch may change the real part of the complex geometric phase by $\pi$ but does not affect the imaginary part.
\end{thebibliography}

\end{document}